\RequirePackage{fix-cm}
\documentclass[smallextended]{svjour3}       
\smartqed  
\usepackage{graphicx}
\usepackage{matlab-prettifier}
\usepackage{amssymb}
\usepackage[english]{babel}
\usepackage[super]{natbib}

\usepackage[letterpaper,top=2cm,bottom=2cm,left=3cm,right=3cm,marginparwidth=1.75cm]{geometry}

\usepackage{amsmath}
\usepackage[colorlinks=true, allcolors=blue]{hyperref}
\usepackage{authblk}
\begin{document}

\title{A fast numerical algorithm for finding all real solutions to a system of N nonlinear equations in a finite domain}

\titlerunning{Finding all real solutions to a system of N nonlinear equations}        

\author{Fernando Chueca-Díez \and
        Alfonso M. Gañán-Calvo 
}


\institute{F. Chueca-Díez \at
              Bristol University, UK \\
              \email{uj21378@bristol.ac.uk}           
           \and
           A. M. Gañán-Calvo \at
              Departamento de Ingeniería Aeroespacial y Mecánica de Fluidos,\\
              Universidad de Sevilla,\\
              41092, Spain; and\\
Laboratory of Engineering for \\
Energy and Environmental Sustainability,\\
Universidad de Sevilla, 41092, Spain\\
\email{amgc@us.es}
}

\date{Received: date / Accepted: date}

\maketitle

\begin{abstract}
A highly recurrent traditional bottleneck in applied mathematics, for which the most popular codes (Mathematica and Matlab) do not offer a solution, is to find all the real solutions of a system of N nonlinear equations in a certain finite domain of the N-dimensional space of variables. We present an algorithm of minimum length and computational weight to solve this problem, resembling a graphical tool of edge detection in an image extended to N dimensions. Once the hypersurfaces (edges) defined by each nonlinear equation have been identified in a single, simultaneous step, the coincidence of the hypersurfaces in the vicinity of all the hyperpoints that constitute the solutions makes the final Newton-Raphson step rapidly convergent to all the solutions with the desired degree of accuracy. As long as N remains smaller than about five, which is often the case for physical systems that depend on fewer than five parameters, this approach demonstrates excellent effectiveness.
\end{abstract}

\section{Introduction}
\label{intro}
Numerical root finding methods such as Newton-Raphson, Broyden's \cite{Broyden1965}, or Levenberg-Marquardt \cite{Levenberg1944} have proven countless times an invaluable tool in the field of engineering. It has permitted the user arrive at a viable solution of a system of equations with relative little computational complexity where finding their analytical solutions becomes, in most cases, an impossible task. A common feature of these numerical methods is that they require an initial set of values which the user must choose to begin the iterative process. The choice of these values can, not only greatly effect the number of iterations required to arrive at an approximate solution (thereby incrementing the computational effort), but will dictate towards which solution the process will converge or if it will converge at all. Sometimes an estimate of the solution is known, either from past experience or from trial and error, but in general this is not the case. Moreover, the system of equations often have a large number of equally important solutions, requiring efficient methods to ensure that all solutions are found in the absence of analytic formulae (e.g. the solution of general polynomial equations of order higher than four, as Galois and Abel independently showed). Without a fail-proof system that guarantees the collection of all solutions in the domain, a search process can become whimsical and unreliable, making the use of an algorithm that can reliably detect all solutions a necessity.

Therefore, the goal is to obtain initial values for all solutions within the selected domain while minimizing the iterative process of the root finding method, with minimal computational and implementation effort. 
Other methods exist to attain this objective\cite{Hsu1984,Alolyan2008,Tsoulos2010,Sidarto2015,Waseem2016,Gao2020,Gong2020}, however, our proposed algorithm has proven to have a substantially lower complexity level in our own experience, resulting in easier implementation and lower computational requirements.

A nonlinear system of equations $F(x)$ may be defined as follows without any loss of generality,

\begin{equation}
    \mathbf{F}(\mathbf{x})=
    \begin{Bmatrix}
        f_1(\mathbf{x}) \\
        f_2(\mathbf{x}) \\
        \vdots \\
        f_n(\mathbf{x})
    \end{Bmatrix}
\end{equation}

with $\mathbf{x} \in S=[a_1,b_1]*[a_2,b_2]\dots[a_n,b_n]\subset \mathbb{R}^n$ and $f_1,f_2,\dots,f_n$ being nonlinear scalar functions such that $f:S\rightarrow \mathbb{R}$. If $\mathbf{x}^*$ is a solution to the system of equations we can say,
\begin{equation}
    \mathbf{F}(x^*)=
    \begin{Bmatrix}
        f_1(\mathbf{x}^*)=0 \\
        f_2(\mathbf{x}^*)=0 \\
        \vdots \\
        f_n(\mathbf{x}^*)=0
    \end{Bmatrix}
\end{equation}
where $\mathbf{x}=\{x_1,...,x_n\}$
Thus, the problem entails discovering the roots of every function in the equation system. Each function defines a (n-1)-dimensional manifold within the n-dimensional domain. Every intersection of {\it all} these manifolds (hyper-surfaces) identifies a solution to the equation system.

\section{Functionality of the algorithm}
\label{sec:1}
In the selected domain, the algorithm generates two distinct sets of discretized spaces. We call these the Stationary set space (SS), and the Perturbed set space (PS). The SS is a fixed N-dimensional grid space encompassing the selected domain. The PS is an N $\times$ N grid consisting of n-displacements of the SS in every direction of the variable $\mathbf{x}$, at an amount of $\Delta x_n = (b_n-a_n)/(N-1)$. Subsequently, the function $\mathbf{F}(\mathbf{x})$ is evaluated at each point of both discretized sets SS and SP.
The Stationary and Perturbed space arrays are stored in two distinct cell arrays, the former being of the form (1,N) and the latter of the form (N,N). Below is a representation of how the cell arrays will appear,\\

\begin{figure}
  \centerline{\includegraphics[width=0.55\textwidth]{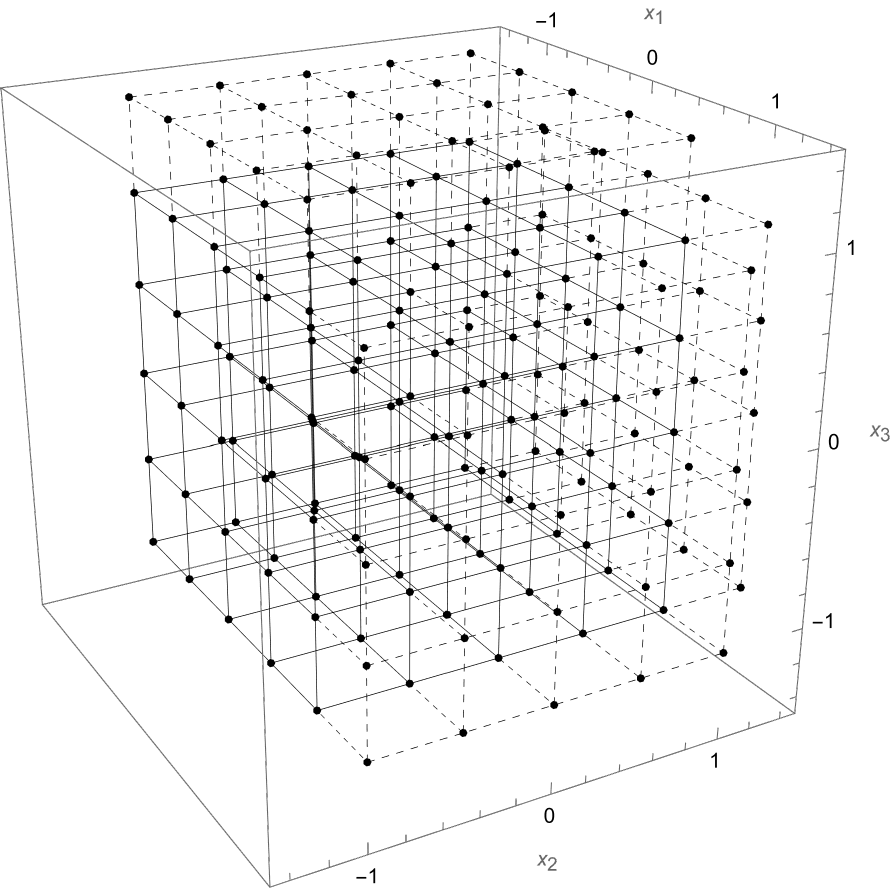}}
  \caption{Illustration of the Stationary and Perturbed discretized spaces (SS and PS, respectively). SS and PS are depicted by continuous lines and dashed lines, respectively, joining the corresponding points of each discretized space.}
\label{f1}
\end{figure}

The SS evaluated by each function can be represented by an array such as:

\begin{equation}
    \mathbf{F}(\mathbf{x}_{N^n})=
    \begin{Bmatrix}
        f_1(\mathbf{x}_{N^n}) \\
        f_2(\mathbf{x}_{N^n}) \\
        \vdots \\
        f_n(\mathbf{x}_{N^n})
    \end{Bmatrix}
\end{equation}
where $\mathbf{x}_{N^n}=\{x_1,...,x_n\}_{N^n}$ represents all possible n-dimensional tuples of the N elements $$x_i\in \left\{a_i,a_i + \Delta x_i,a_i + 2 \Delta x_i,...,a_i + (N-1)\Delta x_i(=b_i)\right\}$$ for $i=1,...,n$, i.e. $\mathbf{x}_{N^n}$ is a super-array of N$^n$ vector values $\mathbf{x}=\{x_1,...,x_n\}$ corresponding to the positions of each point of the SS (see figure \ref{f1}).

The evaluated PS is now written as:
\begin{center}
    \begin{equation}
    \left[\mathbf{F}(\mathbf{x}_{(N^n,1)}),...,\mathbf{F}(\mathbf{x}_{(N^n,n)}) \right] =
    \begin{bmatrix}
f_1(\mathbf{x}_{(N^n,1)}) & f_1(\mathbf{x}_{(N^n,2)}) & \dots & f_1(\mathbf{x}_{(N^n,n)})\\
f_2(\mathbf{x}_{(N^n,1)}) & f_2(\mathbf{x}_{(N^n,2)}) & \dots & f_2(\mathbf{x}_{(N^n,n)})\\
\vdots & \vdots & \ddots & \vdots \\
f_n(\mathbf{x}_{(N^n,n}) & f_n(\mathbf{x}_{(N^n,2)}) & \dots & f_n(\mathbf{x}_{(N^n,n)})\\
\end{bmatrix}
\end{equation}
\end{center}
where $\mathbf{x}_{(N^n,i)}$ is, as before, a super-array of N$^n$ vector values $\{x_1,..,x_i+\Delta x_i,...,x_n\}$ corresponding to the positions of each n-dimensional point of the PS.

The next step is to generate a product array as:
\begin{equation}
    \mathbf{G}=\mathbf{F}(\mathbf{x}_{N^n})\cdot \mathbf{F}(\mathbf{x}_{(N^n,1)})\cdot ... \cdot\mathbf{F}(\mathbf{x}_{(N^n,n)}) =
    \begin{bmatrix}
f_1(\mathbf{x}_{N^n})\cdot f_1(\mathbf{x}_{(N^n,1)}) \cdot f_1(\mathbf{x}_{(N^n,2)}) \cdot ... \cdot f_1(\mathbf{x}_{(N^n,n)})\\
f_2(\mathbf{x}_{N^n})\cdot f_2(\mathbf{x}_{(N^n,1)}) \cdot f_2(\mathbf{x}_{(N^n,2)}) \cdot ... \cdot f_2(\mathbf{x}_{(N^n,n)})\\
\vdots \\
f_n(\mathbf{x}_{N^n})\cdot f_n(\mathbf{x}_{(N^n,n}) \cdot f_n(\mathbf{x}_{(N^n,2)}) \cdot ... \cdot f_n(\mathbf{x}_{(N^n,n)})\\
\end{bmatrix}
\end{equation}
where the product $f_i(\mathbf{x}_{N^n})\cdot \Pi_{k=1}^{n} f_i(\mathbf{x}_{(N^n,k)})$ represents an array of n$\times$N$^n$ values, which are the values of the product of each scalar function $f_i$ at each point of the original SS multiplied by the product of the $n$ values of $f_i$ evaluated at the corresponding (displaced) point of the PS. This operation will reveal each lower vertex of the hyper-cube cells in the SS which contain zeroes of each scalar function $f_i$.

If N becomes a large number, the discrete collection of points corresponding to the negative values of the array $\mathbf{G}$ belongs approximately to each (n-1)-dimensional sheet of the n-dimensional domain where each component of the vector function $\mathbf{F}$ crosses zero.
For example, in a two dimensional space, the points of negative $\mathbf{G}$ values will approach a curve for each scalar $f_1$ and $f_2$. In a three dimensional space, they will approach sheets (or 2-dimensional surfaces).

Thus, all zeros of $\mathbf{F}$ in the selected domain will be defined at those locations where exactly n sheets corresponding to each scalar $f_i$ intersect. These locations are identified, for example, by making all positive values of $\mathbf{G}$ equal to zero, and further multiplying all resulting scalar values of $\mathbf{G}$ at each point. Those points with $\mathbf{G}$ values different from zero will be the approximate locations of the zeros of $\mathbf{F}$.

It is fundamental to perturb the system in all possible directions because we don't know \it a priori \rm the directions of the hyper-sheets $f_n(x_n)=0$ at a given point. For a given function $f(x_n)$, such as $f(x,y)=y \cos (x)$, it may be independent of $y$ around $x=\pi/2$. Therefore, any PS displaced exclusively in the $y-$direction that is performed close to that zero of $f$ will produce no negative $\mathbf{G}$ cells. However, if PS have cells displaced in the $x-$direction, they will capture the zero of $f$ at $x=\pi/2$ since $\mathbf{G}$ will show negative values around that point. In simple words, our method automatically and efficiently performs {\it a one-step, simultaneous local search} at every point within the selected domain defined by the discretisation.

The program subsequently utilizes a root finding method. In the case of the MATLAB code, the standard fsolve() function is utilized and in the case of Mathematica a bespoke Newton-Raphson method has been implemented where the Jacobian can be altered from analytic to numeric. The approximate solutions found by the proposed method are used as initial values to further increase the accuracy of the final solution. Since the initial values are as close to the solution as desired by making 1/N sufficiently small, the root-finding method is extremely fast and efficient.

\section{Results}

The GPU's vector architecture simplifies our initial root location method to a single step, rendering comparisons with other multi-step methods moot. Instead, we will exclusively evaluate its performance using two commonly used codes, Mathematica$\circledR$ and Matlab$\circledR$, which handle matrices differently on CPUs and GPUs.

\subsection{Performance comparison}

For this first subsection, a set of four nonlinear systems of equations where selected to test the effectiveness of the algorithm which has been implemented in MATLAB R2023b and Wolfram Mathematica Version 13.3. All the experiments were preformed using a AMD Ryzen 7 5800HS with Radeon Graphics 3.20 GHz and 16.0 GB of installed RAM. For the given domain of each experiment it was linearly discretized into 500 points, also the experiments where repeated 50 times and the average was recorded.\\

\textbf{Experiment 1: Effati-Nazemi:}

This nonlinear system of equations has been taken from the papers of Effati \cite{Effati2005} where

\begin{equation}
\begin{split}
    f_1(x_1,x_2) & = \cos(2x_1)- \cos(2x_2)-0.4=0\\
    f_2(x_1,x_2) & = 2(x_2-x_1)+ \sin(2x_2)- \sin(2x_1)-1.2=0
\end{split}
\end{equation}

The domain used was $-2\leq x_i\leq 2$, $-10\leq x_i\leq 10$, $-100\leq x_i\leq 100$ and the number of roots inside this domain was unspecified by the literature.\\

\begin{table}[h]
    \centering
    \begin{tabular}{|c|c|c|c|c|}
    \hline
    Domain & Nº Solutions & Time MATLAB(s) & Time Wolf. Mat. & Time Wolf. Mat. \\
    & & &  with Analytical Jac.(s) &  with Numeric Jac.(s) \\
    \hline
    $-2\leq x_i\leq 2$ & 1 & 0.0898 & 0.2161794 & 0.23283381 \\
    $-10\leq x_i\leq 10$ & 13 & 0.1585 & 0.2501860 & 0.25900826 \\
    $-100\leq x_i\leq 100$ & 127 & 0.6391 & 0.3541783 & 0.40448845 \\
    \hline
    \end{tabular}
    \caption{Results for Experiment 1.}
    \label{tab1}
\end{table}

\textbf{Experiment 2: Solution of geometry size of thin wall rectangle
girder section}

From Luo et al \cite{Luo2007}, this is a non-smooth system of equations which gives a singular physical solution to the geometry size of a thin wall rectangle, if we take the non-physical approach as seen in Sidarto \cite{Kuntjoro2015} there is a section of the domain which is undefined in the real set. Therefore, as seen here, the algorithm would fail as it only functions for real functions.

\begin{equation}
    \begin{split}
        f_1(x_1,x_2,x_3)=x_1x_2-(x_1-2x_3)(x_2-2x_3)-165=0\\
        f_2(x_1,x_2,x_3)=\frac{x_1x_2^{3}}{12}-\frac{(x_1-2x_3)(x_2-2x_3)^{3}}{12}-9369=0\\
        f_3(x_1,x_2,x_3)=\frac{2(x_2-x_3)^{2}(x_1-x_3)^{2}x_3}{x_2+x_1-2x_3}-6835=0
    \end{split}
\end{equation}
where $D=\{(x_1,x_2,x_3): -40\leq x_i\leq40 \; i=1,2,3\}$. Their are 6 solutions according to the literature. The three variable system of equations can be simplified into a two variable system of equations rearranging and eliminating $x_2$ \cite{Kuntjoro2015} which is smooth and contained in the real set. Therefore by utilizing this form of the system we are able to arrive at the solutions.\\

\begin{equation}
    \begin{split}
        g_1(x_1,x_3)=\frac{x_1x_2^3}{12}-\frac{(x_1-2x_3)(x_2-2x_3)^3}{12}-9369=0\\
        g_2(x_1,x_3)=\frac{2(x_2-x_3)^2(x_1-x_3)^2x_3}{x_2+x_1-2x_3}-6835=0\\
        x_2=2x_3-x_1+\frac{165}{2x_3}\\
    \end{split}
\end{equation}

\begin{table}[h]
    \centering
    \begin{tabular}{|c|c|c|c|c|}
    \hline
    Domain & Nº Solutions & Time MATLAB(s) & Time Wolf. Mat. & Time Wolf. Mat. \\
    & & &  with Analytical Jac.(s) &  with Numeric Jac.(s) \\
    \hline
    $-40\leq x_i\leq 40$ & 6 & 0.1423 & 1.4845401 & 1.56987439 \\
    \hline
    \end{tabular}
    \caption{Results for Experiment 2}
    \label{tab2}
\end{table}

\textbf{Experiment 3: The reactor problem}

The system of equations below found in Tsoulos et al \cite{Tsoulos2010} ``models two continuous nonadiabatic stirred tank reactors" as described by Floudas \cite{Floudas1999}

\begin{equation}
    \begin{split}
        f_1(x_1,x_2)=(1-R)(\frac{D}{10(1+\beta_1)}-x_1)e^{\frac{10x_1}{1+\frac{10x_1}{\gamma}}}-x_1\\
        f_2(x_1,x_2)=x_1-(1+\beta_2)x_2+(1-R)(\frac{D}{10}-\beta_1x_1-(1+\beta_2)x_2)e^{\frac{10x_2}{1+\frac{10x_2}{\gamma}}}
    \end{split}
\end{equation}
where the parameters $\gamma=1000$, $D=22$, $\beta_i=2$, and R varies from 0.935 to 0.995 with increments of 0.005. $0\leq x_i\leq 1$ and the number of solutions depend on the value of R varying from 1 to 7.\\

\begin{table}[h]
    \centering
    \begin{tabular}{|c|c|c|c|c|}
    \hline
    R & Nº Solutions & Mean Time MATLAB(s) & Time Wolf. Mat. & Time Wolf. Mat. \\
    & & &  with Analytical Jac.(s) &  with Numeric Jac.(s) \\
    \hline
    0.935 & 1 & 0.0802 & 5.8751292 & 5.88312984 \\
    0.940 & 1 & 0.0799 & 8.5863460 & 8.58274849 \\
    0.945 & 3 & 0.0852 & 7.6031261 & 7.50008716 \\
    0.950 & 5 & 0.0891 & 5.4917672 & 5.49120409 \\
    0.955 & 5 & 0.0913 & 8.8077995 & 8.97317449 \\
    0.960 & 7 & 0.0972 & 5.8704372 & 5.91204164 \\
    0.965 & 5 & 0.0908 & 12.2302966 & 12.24315618 \\
    0.970 & 5 & 0.0882 & 9.7278915 & 9.8511479 \\
    0.975 & 5 & 0.0899 & 8.0334215 & 8.09660303 \\
    0.980 & 5 & 0.0903 & 11.7048158 & 12.35137326 \\
    0.985 & 5 & 0.0916 & 7.7216196 & 7.79197757 \\
    0.990 & 1 & 0.0837 & 8.2095828 & 8.33518626 \\
    0.995 & 1 & 0.0809 & 9.8983975 & 10.00050973 \\
    \hline
    \end{tabular}
    \caption{Results form Experiment 3 }
    \label{tab3}
\end{table}

\textbf{Experiment 4: Chen et al.}

From Chen et al \cite{Chen1999} found  in \cite{Kuntjoro2015} the system of equations seen below,

\begin{equation}
    \begin{split}
        f_1(x_1,x_2)=e^{x_1-x_2}- \sin(x_1+x_2)\\
        f_2(x_1,x_2)=x_1^2x_2^2- \cos(x_1+x_2)\\
    \end{split}
\end{equation}

has a domain of $D=-10\leq x_i\leq 10$ and the number of solutions is unspecified by the literature. Figure 1 shows the contour of $F(x_1,x_2)=0$ where the black circles are the solutions found by the algorithm.\\

\begin{table}[h]
    \centering
    \begin{tabular}{|c|c|c|c|c|}
    \hline
    Domain & Nº Solutions & Time MATLAB(s) & Time Wolf. Mat. & Time Wolf. Mat. \\
    & & &  with Analytical Jac.(s) &  with Numeric Jac.(s) \\
    \hline
    $-10\leq x_i\leq 10$ & 6 & 0.051 & 0.3129152 & 0.31447595 \\
    \hline
    \end{tabular}
    \caption{Results for Experiment 4}
    \label{tab4}
\end{table}
\begin{figure}
    \centering
    \includegraphics[width=12.5cm]{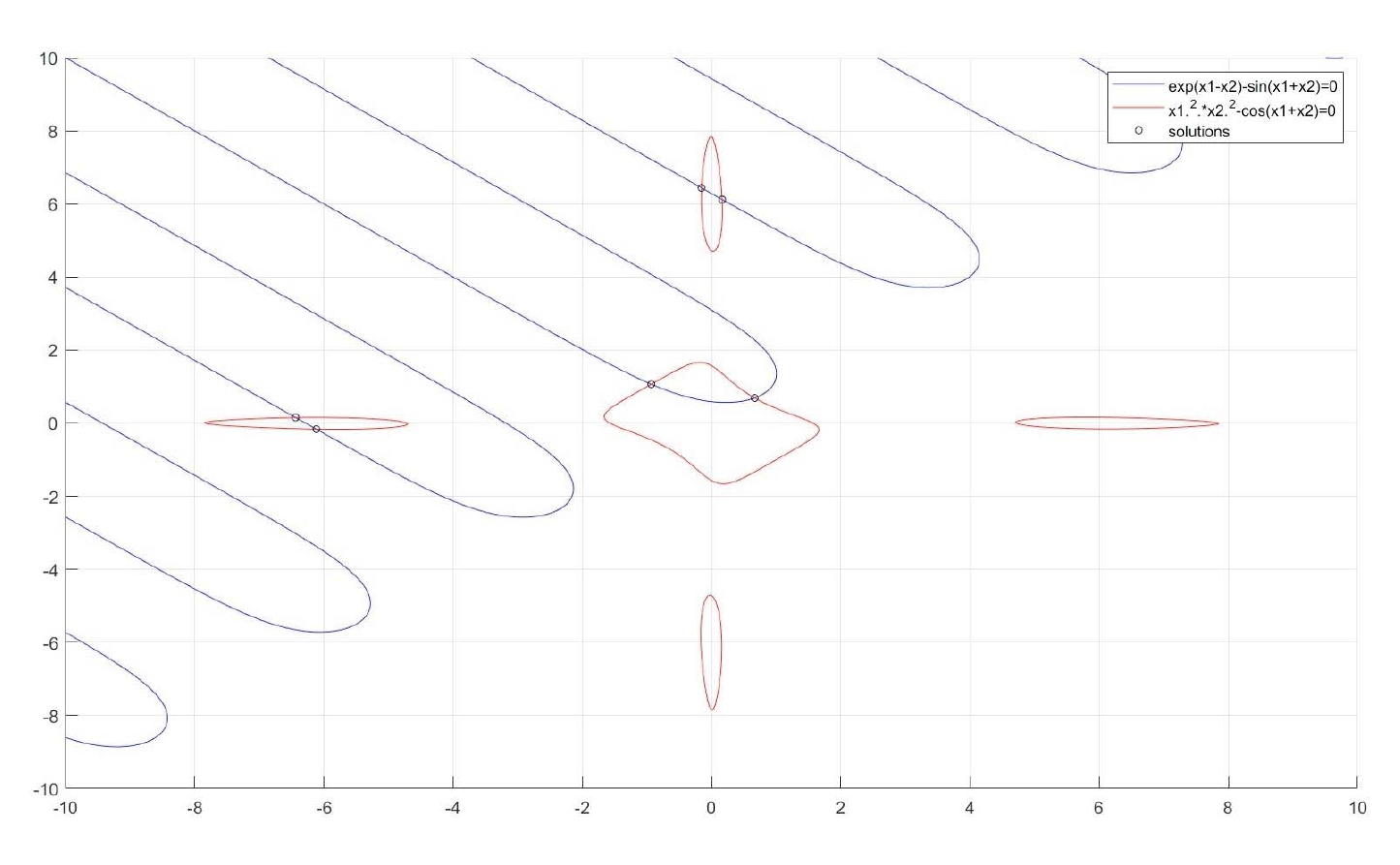}
    \caption{Contour plot of $f_i(x_1,x_2)=0$ for \textbf{Experiment 4} where the black circles are the solutions arrived at utilizing the algorithm}
    \label{f2}
\end{figure}

We can quickly draw some insightful conclusions from the results of these experiments. At first glance we can see the MATLAB version of the algorithm is in almost every case a factor of some 100 times faster than the one in Mathematica. It can also be seen that MATLAB is much more consistent and is not affected by the form of the functions. Looking more in detail in Experiment 1 it can be seen that for the largest domain,  $-100\leq x_i\leq 100$, Mathematica is faster. It is believed this is because, as stated before, the algorithm implemented in MATLAB utilizes the function fsolve() to achieve final refinement which utilizes the  Levenberg-Marquardt method to reach the final solution whereas the algorithm in Wolfram Mathematica uses a bespoke Newton-Raphson method. It can also be the case that this area of the code in MATLAB hasn't been optimised to its maximum extent and is creating a bottle neck, this is something that could be revised in future work.

As stated at the start of the section the level of discretisation of the domain was maintained constant, independently of the size of the domain. This was done so that the results between different experiments could be compered as fairly as possible. To make sure we obtained all the solutions, regardless of how often the systems of equations produced them, we selected a very fine mesh. In the following subsection it will be explored how the level of discretisation and number of variables affects the run-time of the algorithm.

\subsection{Discussion}

Because of the nature of the algorithm the level of discretisation required is significantly low, achieving accurate results with as little as 20 points in each dimension for a domain of $-10 \leq x_1, x_2 \leq 10$. This can be attributed to the fact that close to $F(x_i)=0$ multiple approximate solutions appear (due to the multiplicity of local search directions provided by the algorithm) which are all used by the final refinement method to arrive at $F(x_i) \rightarrow 0$. In most cases all of these will tend to the same solution, but if two or more solutions are close together these close but different multiple starting points would lead to the differentiated roots with a rather coarse mesh (see figure \ref{f2}).

Nonetheless, it was important to know how the run-time would be affected for a different number of points and different number of dimensions. For this a similar approach was taken as in the first section. Keeping the domain constant $ -10 \leq x_i \leq 10$ the number of point where increased for every variable in steps of 10 starting from a minimum of 20 points and making the maximum 1000, it will be seen that in most cases 1000 wasn't reached. Each step was repeated 50 times and the mean was calculated. The experiment was preformed only on the MATLAB code as from the previous section it was seen that runt-time of the Mathematica code, both  with the numeric and analytic Jacobian, where greatly affected by the type of function that was being analysed.

\begin{figure}[h]
    \centering
    \includegraphics[width=\linewidth]{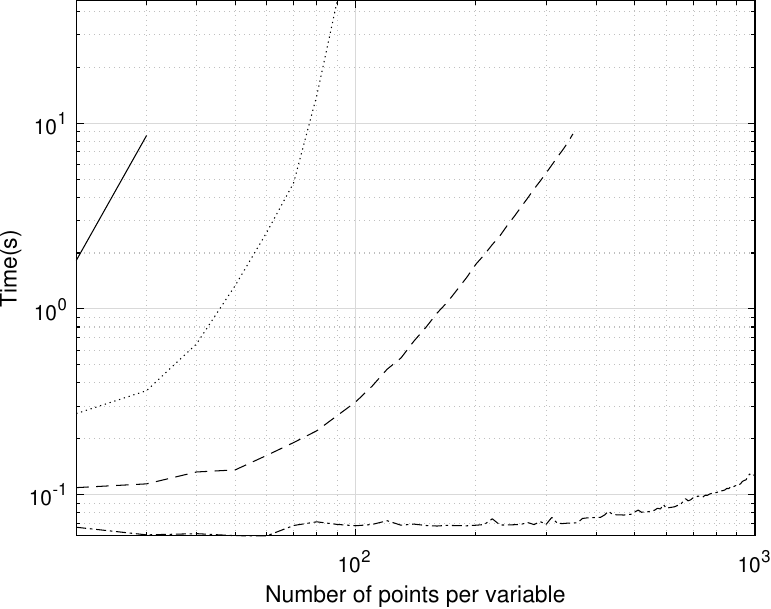}
    \caption{Results for Time(s) with respect to the number of points per variable where 2D:'-.', 3D:'--', 4D:'dotted', 5D:'continuous'.}
    \label{f3}
\end{figure}

In figure \ref{f3} we can observe the main drawback to this method, since the tensors created to evaluate the functions at the different points are of rank n, where n is equal to the number of dimensions the system of equation, this means the computational effort and memory required scales proportional to $N^n$ where $N$ is the number of discretized points per variable assuming they are equal for each. Because of this, it was only possible to go up to N$=1000$ points with a 2 dimensional system, the other points having to be cut short, either because of not sufficient memory or prolonged run-time.

This can be argued to be a fatal flaw of the method, and although for systems of large dimensions this is true, for systems of up to 4 dimensions the algorithm is extremely efficient, as we must recall the discretisation of the system can be made extremely coarse and still obtain accurate results. In this regard, most physical systems are likely to have no more than four independent variables that affect them to a comparable degree. Hence, this approach can be particularly appropriate for systems with a few degrees of freedom or systems of large numbers, such as continuous media, economy, social or ecology models which can be represented by just a few parameters over a broad range where the solution may reside.

\begin{acknowledgements}
AMGC is partially supported by the Ministry of Science and Innovation, grant no. PID2019-108278RB-C31.
\end{acknowledgements}

%
\section*{Author contributions}
FCD: paper writing, Matlab and Mathematica codes, analysis. AMGC: project, algorithm concept, Mathematica code, analysis, writing and supervision.
%
%
\section*{Conflict of interests}
The authors declare that they have no conflict of interest.

\appendix

\section{Appendix: Required inputs by the user}

For the code to arrive at the desired solution it is required by the user to specify the system of equations, the domain the user is interested in, and the level of discretization. The code makes use of anonymous functions to describe the system of equations by creating an array in text form where each equation can be written. Then, by using str2func inside a for loop with N iterations, a (1,N) cell array is constructed where each cell stores a function. The domain and level of discretization is then selected by stating the region of each variable and the number of linearly spaced intervals desired; this has the benefit that each variable is independent of each other and therefore if it is known that a singular variable has a very high frequency of ``solution production" the spacing of intervals can be decreased without hampering run-times significantly. A 2-D example of how the input displays in the code is seen below:\\

\begin{lstlisting}[style=Matlab-editor]
%Non-linear system of equations
equations = {'@(x1,x2)x1.*cos(0.5*x2)','@(x1,x2)-x1+0.5.*x2.^2'};%Set your equations
N = numel(equations);
f = cell(1,N);

%Creation of domain
x1=linspace(-10,10,11); x2=linspace(-10,10,11); %Set your domain
 (number of variables must be equal to number of equations)
domain = {x1,x2};
for i = 1:N
    f{i} = str2func(equations{i});
end
\end{lstlisting}

\section{Appendix: Line by line description of how the code works}

Lines which description is empty does not necessary mean the lines are completely blank as they will contain comments or notes. By empty, it means that they contain no value to the code and have no effect on it.

Line 1-2: Defining of system of equation as seen above.

Line 3: Utilizing numel(equations) to find the number of elements in the array and subsequently the dimension of the system.

Line 4: Producing an empty (1,N) cell array to later store each of the anonymous functions.

Line 5-6:Empty.

Line 7: Defining the domain of each variable, as stated before each variable is independent of each other and therefore the user can define them as it best suits them.

Line 8: Creating a cell array named domain, each cell contains the discretized domain of each variable.

Line 9: Creation of empty cell arrays of length (1,N) named domain1 and domain2 to be used subsequently.

Line 11-15, for loop i=1:N:

Line 12: Utilizing str2func to convert the text representation of each cell in equations to a function handle which is stored in a subsequent cell in the cell array f.

Line 13-14: Producing from domain a ``stationary" domain and a ``moved" domain. This is done by removing the final value of domain to produce domain1 and removing the starting value to produce domain2, producing the deviation.

Line 16-17: Empty.

Line 18: Creating (2,N) empty cell array named X\_1 which will be used subsequently.

Line 19-20: Utilizing ndgrid to produce an n-dimensional full grid of domain1 and domain2. X\_1\{1,:\} is the Stationary space and X\_2\{1,:\} is the "displaced Stationary" space which will subsequently be used to produce the Perturbed set space.

Line 21: Creating (1,N) empty cell array named X\_M which will be used subsequently.

Line 22-30, for loop i=1:N:

Line 23: Creating (1,N) empty cell array named inner\_cell

Line 24-27, nested for loop j=1:N:

Line 25: The Stationary space, X\_1\{1,:\} fills all the cells in inner\_cell

Line 26: The displaced variable, fills only cell i in the array. Then this cell array is stored in one of the cells of X\_M in Line 28. With each iteration of the loop the variable that is displaced changes, thereby creating the Perturbed set space.

Line 28: Functionality of this line of code explained in Line 26.

Line 30-31: Empty.

Line 32: Creation empty cell array F (1,N) and F\_2 (N,N).

Line 33-38, for loop i=1:N:

Line 34: Evaluating every function in the Stationary space and storing the output of the N dimensional array in one cell in F as can be seen in (1).

Line 35-37, nested for loop j=1:N:

Line 36: Each function is evaluated N times and each time it is evaluated by one of the Perturbed spaces, resulting in (2).

Line 39-41: Empty.

Line 42: Creation empty cell array F\_pr (N,N).

Line 43-48, for loop i=1:N:

Line 44-47, nested for loop j=1:N:

Line 45: Element wise multiplication is utilized between function N in the Stationary space and the set of Perturbed spaces. Allocating each of the arrays produced to one of the cells of F\_pr.

Line 46: All the positive value elements are converted to zero.

Line 49-50: Empty

Line 51: Creation empty cell array F\_pr\_sum (N,N)

Line 52-57, for loop i=1:N:

Line 53: The empty cell, \{1,i\} from F\_pr\_sum is equated to cell \{i,1\} from F\_pr; this is done so that, subsequently, the summation of the different arrays is compatible, as now the arrays will be of equal size.

Line 54-56, nested for loop j= 1:N-1:

Line 55: Each cell of every row of F\_pr creating a single array, which is then stored in one of the cells of F\_pr\_sum.

Line 58-59: Empty.

Line 60: Array inter, of equal size to a cell in F\_pr\_sum is constructed. Therefore, to shorten code this array is simply equated to F\_pr\_sum\{1,1\}.

Line 61-63, for loop i=1:N-1:

Line 62: Element wise multiplication is utilized to find the point of intersection between all function as these will be the only nonzero elements in inter, an N-dimensional array.

Line 64: Empty.

Line 65: MATLAB find function is utilized to produced a column vector containing the linear indices of each nonzero element in array inter.

Line 66: Creation of two empty cell array of size (1,N) named X\_sol and X\_pos.

Line 67: ind2sub converts the linear indices stored in idx to equivalent multidimensional subscripts of equivalent size to size(inter) (MATLAB Help Centre). Therefore, in X\_pos we store the indice value not the value itself, where each cell of X\_pos has a column vector of the solutions of variable i.

Line 68: Empty.

Line 69-71, for loop i=1:N:

Line 70: X\_pos is utilized to obtain the approximate solutions by applying the indices values to the given domain of variable i, thereby obtaining an array of the approximate solutions, X\_sol.

Line 72-73: Empty.

Line 74: Create a zero array of row size equal to the number of columns of X\_sol, and column size N.

Line 75: Create a symbolic system of equations, unlike the start where we wanted to have them separated for independent use.

Line 76: Converting the symbolic expression of the system of equations into a single function handle.

Line 77: Creating a (1,N) zero array named initial\_pos.

Line 78-90, for loop i=1:size(((X\_sol\{1,1\})'),1):

Line 79-81, nested for loop j=1:N:

Line 80: Create a (1,N) array of initial values which with every loop iteration it cycles through all the set of approximate solutions found above.

Line 82: Utilizing fsolve to arrive at a solution to the system of equations from the initial values.  The variable named solution stores the output of fsolve, fval is the final output of the function, and exitflag are integer values which correspond to reasons the iterations stopped.

Line 83-89, if statement exitflag<1: If the value of exitflag is less than 1 it means the algorithm has not arrived at a solution and stopped. For further information proceed to the documentation of fsolve in MATLAB Help Centre.

Line 84: row i of arr\_solution is converted to NaN.

Line 86-88, nested for loop j=1:N:

Line 87: Values of array named solution are stored in row i of array arr\_solution.

Line 91: Values in arr\_solution are rounded to six decimal places and all repeating rows are eliminated using the unique function.

Line 92: All values NaN are eliminated.



\end{document}